# GEODETIC GRAPHS HOMEOMORPHIC TO A GIVEN GEODETIC GRAPH

Carlos E. Frasser, George N. Vostrov

*Department of Applied Mathematics, Odessa National Polytechnic University, Ukraine*

This paper describes a new approach to the problem of generating the class of all geodetic graphs homeomorphic to a given geodetic one. An algorithmic procedure is elaborated to carry out a systematic finding of such a class of graphs. As a result, the enumeration of the class of geodetic graphs homeomorphic to certain Moore graphs has been performed.

## 1. Introduction.

If $G$ is a *graph*, then $V(G)$ and $E(G)$ denote its *vertex set* and *edge set*, respectively. In this paper a graph $G$ is connected, undirected, and without loops or multiple edges. A *path P* from $v_0$ to $v_n$ is a sequence $v_0v_1,\ldots,v_n$ of different successive vertices which are joined by an edge. A *circuit C* is a sequence $v_0v_1,\ldots,v_nv_0$ of different successive vertices which are also joined by an edge. The *length* of $P = v_0v_1,\ldots,v_n$ is the number of edges it contains and will be denoted either by $|P| = |v_0v_1,\ldots,v_n|$ or by $|e_1e_2,\ldots,e_n|$, where $e_1, e_2,\ldots,e_n$ is a sequence of edges that join vertices $v_0,v_1,\ldots,v_n$. A circuit $C$ is *even* or *odd* if its length is even or odd, respectively. We can assign an orientation to each circuit $C$. If $C$ is an even circuit and $u, v \in V(C)$, we say that $u, v$ are $C$-opposite if $|C(u, v)| = |C(v, u)|$.

The distance between $u, v \in V(G)$, denoted $d_G(u, v)$, is the length of a shortest path connecting these vertices. The diameter of $G$, denoted $d(G)$, is the greatest distance between any pair of its vertices. If $u$ and $v$ are two different vertices of $G$, a *geodesic* $\Gamma(u, v)$ in $G$, is a path of shortest length whose endpoints are $u$ and $v$. Clearly, $d_G(u, v) = |\Gamma(u, v)|$. A Graph $G$ is *geodetic* [6] if, given any vertices $u$ and $v$ of $G$, there exists a unique geodesic $\Gamma(u, v)$ in $G$.

The degree of a vertex $v$, denoted $deg(v)$, is the number of edges incident to $v$. $G$ is said to be *regular* of degree $k$ if every vertex of $G$ has the same degree $k$. A *node v* is defined as a vertex for which $deg(v) \geq 3$. A *segment* is a path whose only nodes are its endpoints $u$ and $v$ and it is denoted $S(u, v)$. The girth of a graph $G$, denoted $g(G)$, is the length of a shortest circuit contained in $G$.

Two graphs $G_1$ and $G_2$ are said to be isomorphic, denoted $G_1 \cong G_2$, if there exists a one-to-one function from $V(G_1)$ onto $V(G_2)$ that preserves adjacency. Two graphs are said to be homeomorphic if and only if each can be obtained from the same graph by insertion of vertices onto the edges of one or both graphs. Notice that every graph is homeomorphic to itself. A complete graph is a graph in which every pair of vertices is adjacent. The complete graph with $n$ vertices is denoted $K_n$. A block is a graph with vertex connectivity $> 1$.

If $H$ is a subgraph of $G$, $G - H$ denotes the subgraph of $G$ obtained by deleting $V(H)$ from $V(G)$ and removing all edges from $G$ that have an endpoint in $V(H)$. A subset $S$ of $V(G)$ is said to *generate* a subgraph if $H$ is the section subgraph on $S$; that is, $V(H) = S$ and $H$ contains all edges of $G$ connecting two vertices of $S$ [11]. A *clique* is defined as a maximal complete subgraph $K_n$, $n \geq 3$; that is, a complete subgraph on at least three vertices which is contained in no larger complete subgraph.

A regular graph *G* of degree *k* and diameter *d* is called a *Moore graph* if

$$|V(G)| = 1 + k \sum_{i=1}^{d} (k-1)^{i-1}$$

For a Moore graph of diameter 2, this condition becomes $|V(G)| = 1 + k^2$.

From Stemple [11] and Hoffman & Singleton [4], we have

**Theorem 1.** Let *G* be a geodetic block of diameter 2. If *G* does not contain cliques $K_n$, $n \geq 3$, then *G* is a Moore graph, where $k = 2, 3, 7, 57$. If $k = 2$, then *G* is a circuit of length 5. If $k = 3$, then *G* is the Petersen graph. If $k = 7$, then *G* is the Hoffman-Singleton graph.

The existence of a graph as described in the previous theorem with $k = 57$ is still undecided.
Now, we will examine some results related to geodetic graphs which are formulated in terms of the general ideas previously exposed.
The following characterization was established by Stemple and Watkins [13].

**Lemma 1.** A connected graph *G* is geodetic if and only if *G* does not contain an even circuit *C* such that for every pair of *C*-opposite vertices *u*, *v* $d_G(u, v) = d(C)$.

From Stemple [12], we have

**Theorem 2.** Let *G* be homeomorphic to $K_4$; *G* is geodetic if and only if:
   (1) The six segments of *G* are geodesics.
   (2) Each circuit of *G* that contains exactly three segments is odd.
   (3) All circuits of G that contain exactly four segments have equal length.

A graph *G* homeomorphic to $K_4$ which satisfies Theorem 2's conditions (1), (2), and (3) is shown in Figure 1.

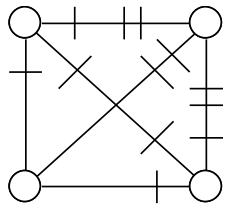

Fig. 1. A geodetic graph homeomorphic to $K_4$.

We define a $K_n^i$ graph as one that is homeomorphic to a complete graph $K_n$ with nodes $u_r$ ($r = 1, 2, \ldots, n$) together with a function *i* which assigns non-negative integers $i(u_r)$ to the nodes such that, given two nodes $u_r$ and $u_s$ of $K_n^i$, $|S(u_r, u_s)| = i(u_r) + 1 + i(u_s)$.

We will sometimes refer to a $K_n^i$ as a Plesnik graph and to the $i(u_r)$ as Plesnik numbers [1]. From Plesnik [9] and Stemple [12], we have

**Theorem 3.** A graph *G* homeomorphic to a complete graph $K_n$ is geodetic if and only if it is a $K_n^i$ graph (see Figure 2).

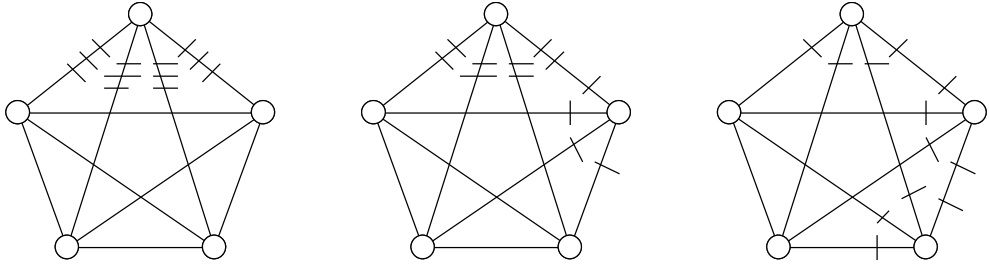

Fig. 2. A set of non-isomorphic geodetic graphs $K_5^3$ homeomorphic to a complete graph $K_5$.

Over the years, several authors have described clever techniques of generating different classes of geodetic graphs homeomorphic to a given geodetic one [5, 9, 10]. Trying to solve the problem, some particular families of these classes of graphs have been completely characterized [7, 12]. However, there is no a general procedure that leads to the construction of the class of all geodetic graphs homeomorphic to a given geodetic one. In the following section, we attempt to describe a technique for solving this fundamental problem, which can be extended to any geodetic graph.

## 2. The Problem of Generating the Class of all Geodetic Graphs Homeomorphic to a Given Geodetic Graph.

We already mentioned that a Moore graph of type $(k, d)$ is defined as a regular graph of degree $k$, diameter $d$, and girth $g = 2d + 1$. For any $k$, there exists a Moore graph of type $(k, 1)$, represented by a complete graph of degree $k$, and only one Moore graph of type $(3, 2)$, represented by the Petersen graph. Every Moore graph of type $(k, d)$ may be represented as the union of a tree $T = T(k, d)$ with levels $0, 1, 2,…, d$ (see Figure 3) and a set $S$ of edges which suitably connect the vertices of level $d$ ($S$ can be here considered as a subgraph of degree $k - 1$). Any vertex of the graph may be used as origin or topmost point.

The described procedure to generate a Moore graph and the following theorem are taken from [2].

**Theorem 4.** Every circuit in a Moore graph $G$ of type $(k, d)$ has length no less than $2d + 1$.

Let $G$ be a graph and let $T$ be a spanning tree of $G$; that is, a subgraph of $G$ which is a tree containing all vertices of $G$. A circuit is called *fundamental* if it is a circuit created by adding in $T$ an edge $e$ of $G$ that is not in $T$. Each of the fundamental circuits is linearly independent from the remaining circuits because it includes an edge $e$ that does not belong to any other fundamental circuit. So, the set of fundamental circuits forms a basis for the circuit space. A circuit basis constructed this way is called a *fundamental circuit basis*.
In this direction, we have established

**Theorem 5.** Every Moore graph $G$ of type $(k, d)$ has a fundamental circuit basis composed of odd circuits of minimal length $2d + 1$.

**Proof.** Every Moore graph of type $(k, d)$ may be represented as the union of a spanning tree $T = T(k, d)$ with levels $0, 1, 2,…, d$ and a set $S$ of edges which suitably connect the vertices of level $d$. Therefore, any edge of $S$ that is added to $T = T(k, d)$ generates in $T$ a circuit $C$ containing $2d$ edges of $T$; that is, each edge of $S$ generates a fundamental circuit of length $2d + 1$ which, according to Theorem 4, is minimal. That way, we are able to construct a fundamental circuit basis composed of odd circuits of minimal length $2d + 1$ and so the Theorem follows.

The Theorem on the general number of different circuits of lengths $(2d + 1)$ and $(2d + 2)$ contained in a Moore graph was proved by Friedman [2].

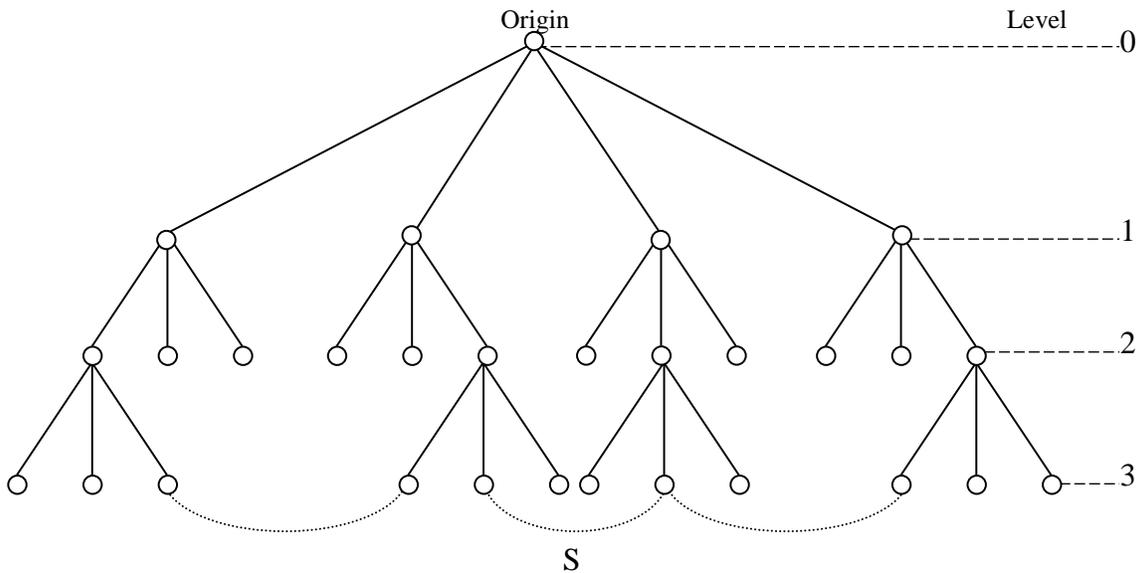

Fig. 3. Tree $T(4, 3)$ and some edges of $S$.

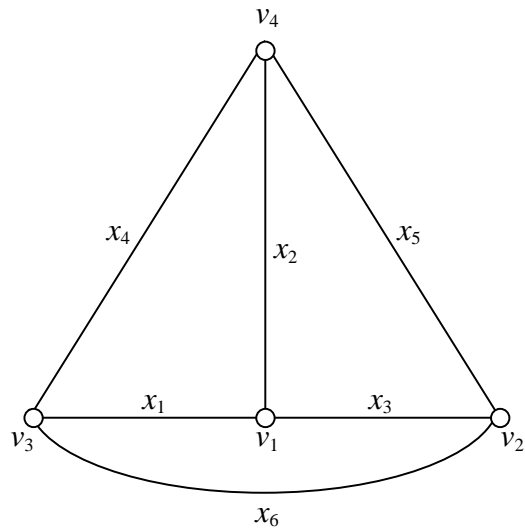

Fig. 4. A complete graph $K_4$ represented in the form of a tree $T(3, 1)$ and all edges of $S$.

**Theorem 6.** The number of distinct circuits of length $(2d + 1)$ and $(2d + 2)$ in a Moore graph of type $(k, d)$ is, respectively,

(i)  $k(k-1)^d[k(k-1)^d - 2]/[2(2d+1)(k-2)]$,

(ii) $k(k-1)^d[k(k-1)^d - 2]/[2(2d+2)]$.

Let $x_1, x_2, x_3, x_4, x_5, x_6$ be the edges of a complete graph $K_4$. Let us construct a system of six linear equations whose unknowns are the edges $x_1, x_2, x_3, x_4, x_5, x_6$ of $K_4$ (see Figure 4) as follows:

$$\begin{aligned} x_1 + x_2 \quad\quad + x_4 \quad\quad\quad\quad &= 2k_1 + 1 \\ x_2 + x_3 \quad\quad + x_5 \quad\quad &= 2k_2 + 1 \\ x_4 + x_5 + x_6 &= 2k_3 + 1 \\ x_1 + x_2 \quad\quad\quad\quad + x_5 + x_6 &= 2d + 2 \\ x_2 + x_3 + x_4 \quad\quad + x_6 &= 2d + 2 \\ x_1 + \quad\quad x_3 + x_4 + x_5 \quad\quad &= 2d + 2 \end{aligned} \quad (1)$$

where $d$ and $k_j$, $j = 1, 2, 3$ are positive integers such that $k_j \leq d$, $j = 1, 2, 3$.

Because we are interested only in the solutions in natural numbers of linear system (1) and $K_4$ is geodetic, we have defined system (1) as the *geodetic system of Diophantine equations* associated with $K_4$.

We say that a natural solution $(x_1, x_2,\ldots, x_6) = (a_1, a_2,\ldots, a_6)$ of system (1) determines a graph $G$ homeomorphic to $K_4$ if such a solution determines the length of the segments of $G$; that is, the values of unknowns $x_1, x_2,\ldots, x_6$ determine, in that order, the length of segments $x_1, x_2,\ldots, x_6$. Conversely, a graph $G$ homeomorphic to $K_4$ determines a natural solution of system (1) if the values of the length of the segments of $G$ (which means, the values of unknowns $x_1, x_2,\ldots, x_6$) satisfy the equations of system (1). Notice that in system (1), the value of $d$ represents the diameter of a given graph $G$ homeomorphic to $K_4$.

It is easy to realize that the unknowns of the first three equations of system (1) are the edges of the circuits belonging to a fundamental circuit basis associated with a graph $K_4$. This basis is indeed composed of odd circuits of minimal length 3 with respect to a spanning tree $T = T(3, 1)$. On the other hand, the unknowns of the last three equations of system (1) determine the edges of all circuits of length 4 contained in $K_4$. Notice that according to Theorem 6, $K_4$ has four circuits of length 3 and three of length 4.

For a graph $G$ homeomorphic to a complete graph $K_4$ and its geodetic system of Diophantine equations, we have established

**Theorem 7.** Every natural solution of the geodetic system of Diophantine equations associated with a complete graph $K_4$ determines a geodetic graph homeomorphic to $K_4$ and every geodetic graph homeomorphic to $K_4$ determines a natural solution of $K_4$'s geodetic system of Diophantine equations.

**Proof:** We will prove the first part of the theorem by contradiction. Let us assume that being $k_j \leq d$, $j = 1, 2, 3$ for fixed positive integer values of $d$ and $k_j$, $j = 1, 2, 3$ system (1) has a natural solution that generates a non-geodetic graph $G$ homeomorphic to $K_4$. From the construction of system (1) it follows that $G$ satisfies Theorem 2's conditions (3) because the equations that define all circuits of length 4 are equal to the same magnitude $2d + 2$; that is, the circuits containing four segments are all equal in length. In fact, Theorem 2's conditions (2) are also satisfied. It is evident that the equation defining the circuit of length 3 that is not included in system (1) also determines and odd number:

$$x_1 + 2x_2 + x_3 + 2x_4 + 2x_5 + x_6 = 2(k_1 + k_2 + k_3) + 3$$
$$2(x_2 + x_4 + x_5) + (x_1 + x_3 + x_6) = 2(k_1 + k_2 + k_3) + 3$$
$$(x_1 + x_3 + x_6) = 2[(k_1 + k_2 + k_3) - (x_2 + x_4 + x_5)] + 3.$$

So, each circuit containing three segments is odd. Therefore, it must happen that $G$ does not satisfy at least one of Theorem 2's conditions (1). Assume that $(v_3, v_1)$-segment is not a geodesic in $G$. Then a geodesic $\Gamma(v_3, v_1)$ in $G$ must be one of the following paths (see Figure 4):

(a) $v_3v_4v_1$, where $|v_3v_4v_1| = x_2 + x_4$,
(b) $v_3v_2v_4v_1$, where $|v_3v_2v_4v_1| = x_2 + x_5 + x_6$,
(c) $v_3v_4v_2v_1$, where $|v_3v_4v_2v_1| = x_3 + x_4 + x_5$,
(d) $v_3v_2v_1$, where $|v_3v_2v_1| = x_3 + x_6$.

Now, assume that path $P = v_3v_4v_1$ is a geodesic $\Gamma(v_3, v_1)$ joining vertices $v_3$ and $v_1$ in $G$. Then the following four inequalities take place

$$x_1 > x_2 + x_4 \quad \text{which means} \quad x_1 > x_2 + x_4 \quad (2)$$
$$x_2 + x_5 + x_6 > x_2 + x_4 \quad\quad\quad\quad\quad\quad x_5 + x_6 > x_4 \quad (3)$$
$$x_3 + x_4 + x_5 > x_2 + x_4 \quad\quad\quad\quad\quad\quad x_3 + x_5 > x_2 \quad (4)$$
$$x_3 + x_6 > x_2 + x_4 \quad\quad\quad\quad\quad\quad x_3 + x_6 > x_2 + x_4 \quad (5)$$

Adding (3), (4), and (5), we obtain

$$x_3 + x_5 + x_6 > x_2 + x_4,$$

and so
$$x_3 + (2d + 1) \geq x_3 + (x_4 + x_5 + x_6) > x_2 + 2x_4,$$
$$x_3 + (2d + 1) > x_2 + 2x_4,$$
$$x_3 + 2d \geq x_2 + 2x_4,$$
$$x_1 + x_3 - x_4 + 2d \geq x_1 + x_2 + x_4.$$

Therefore,
$$x_1 + x_3 - x_4 = 1,$$
$$x_1 = 1 - x_3 + x_4. \quad (6)$$

Substituting (6) into (2), we finally obtain:
$$x_2 + x_3 < 1.$$

So, the addition of the length of two distinct segments in $G$ is less than 1. This is a contradiction. If we assume that any of the other paths mentioned above that are different of $(v_3, v_1)$-segment is also a geodesic in $G$, we will arrive at a similar contradiction. Thus, $(v_3, v_1)$-segment is a geodesic in $G$. In a similar fashion, it can be proved that any other segment of $G$ different of $(v_3, v_1)$-segment is also a geodesic in $G$. That way, we proved that $G$, which, according to our assumption, is non-geodetic, actually satisfies Theorem 2's conditions (1), (2), and (3). But again, this is a contradiction. Therefore, system (1) generates natural solutions that determine only geodetic graphs homeomorphic to $K_4$.

We will prove the second part also by contradiction. Let us assume that there exists at least one geodetic graph $G$ homeomorphic to $K_4$ that determines a natural solution which does not satisfy system (1). Then this solution does not satisfy at least one of the six equations of system (1); that is, $G$ determines a natural solution that does not satisfy either one of Theorem 2's conditions (2) or one of Theorem 2's conditions (3). But according to Theorem 2, this contradicts the fact that G is geodetic. Thus, Theorem 7 follows.

**Corollary 1.** The set of all natural solutions of the geodetic system of Diophantine equations associated with $K_4$ determine the class of all geodetic graphs homeomorphic to $K_4$.

In Section 3 of this research is described an algorithm that allows us to generate the set of all natural solutions of the geodetic system of Diophantine equations, which can be applied not only to the system associated with $K_4$, but to the system of any other geodetic graph. In addition, the general number of natural solutions of the geodetic system of Diophantine equations associated with certain Moore graphs has been calculated.

Now, a result similar to that one established in Theorem 7 can be extended to geodetic graphs homeomorphic to a complete graph $K_n$.

Let $H$ be a subgraph of a given graph $G$. Let us assume that $H$ is homeomorphic to $K_4$. It is said that $H$ is a $K_4^H \cdot$ subgraph of $G$ if each segment of subgraph $H$ is also a segment of $G$; that is, $H$ contains exactly four nodes of $G$ and a segment of $G$ joining each pair of these nodes.

In this direction, Stemple [12] established

**Theorem 8.** Let $G$ be a graph homeomorphic to a complete graph $K_n$. $G$ is geodetic if and only if every $K_4^H$ - subgraph of $G$ is geodetic.

Let $v_1,...,v_n$ be the vertices of $K_n$ and let $x_1,...,x_{n(n-1)/2}$ be the edges of $K_n$. Let us construct a system of $6[(n-1)^2(n-2)^2 - 2(n-1)(n-2)]/24 = [(n-1)^2(n-2)^2 - 2(n-1)(n-2)]/4$ Diophantine linear equations whose unknowns are $x_1,...,x_{n(n-1)/2}$ as follows.

Firstly, we consider one $K_4^H$ - subgraph of the $[(n-1)^2(n-2)^2 - 2(n-1)(n-2)]/24$ $K_4^H$ - subgraphs of a complete graph $K_n$. Then we construct the corresponding system (1) of six Diophantine linear equations. Now, we repeat the procedure for the remaining $K_4^H$ - subgraphs until we obtain a system of $[(n-1)^2(n-2)^2 - 2(n-1)(n-2)]/4$ equations. This new system is called the geodetic system of Diophantine equations associated with a complete graph $K_n$.

The constant terms are designated so that for fixed positive integer numbers $d_m$, $m = 1, 2,...,[(n-1)^2(n-2)^2 - 2(n-1)(n-2)]/24$, there are positive integer numbers $k_j$, $j = 1,2,...,[(n-1)^2(n-2)^2 - 2(n-1)(n-2)]/8$ such that $k_j \leq d_m$ for all $j = 3m - 2, 3m - 1, 3m$, $m = 1, 2,...,[(n-1)^2(n-2)^2 - 2(n-1)(n-2)]/24$. Notice that the maximum value chosen among those $d_m$, $m = 1, 2,...,[(n-1)^2(n-2)^2 - 2(n-1)(n-2)]/24$ is the diameter of $G$.

**Theorem 9.** Every natural solution of the geodetic system of Diophantine equations associated with a complete graph $K_n$ determines a geodetic graph homeomorphic to $K_n$ and every geodetic graph homeomorphic to $K_n$ determines a natural solution of $K_n$'s geodetic system of Diophantine equations.

**Proof:** Let us prove the first part of the theorem. Assume that the geodetic system of Diophantine equations associated with $K_n$ has a natural solution $(x_1,...,x_{n(n-1)/2}) = (a_1,...,a_{n(n-1)/2})$ that determines a graph $G$ homeomorphic to $K_n$. Since this solution satisfies the geodetic subsystem of Diophantine equations of each $K_4^H$ - subgraph of $K_n$ and according to Theorem 7, the natural solutions of each of these subsystems determine only geodetic graphs homeomorphic to $K_4$, then according to Theorem 8, $G$ must be geodetic.

Now, we will prove the second part of the theorem by contradiction. Assume that there exists at least one geodetic graph $G$ homeomorphic to $K_n$ that determines a natural solution which does not satisfy the geodetic system of Diophantine equations associated with $K_n$. Then this solution does not satisfy at least one of the equations of this system and therefore, one of the equations of some geodetic subsystem of Diophantine equations of the $K_4^H$ subgraphs of $K_n$; that is, $G$ contains a $K_4^H$-subgraph that is non-geodetic. But according to Theorem 8, this contradicts the geodeticity of $G$.

**Corollary 2.** The set of all natural solutions of the geodetic system of Diophantine equations associated with $K_n$ determine the class of all geodetic graphs homeomorphic to $K_n$.

Because of the way a Moore graph is constructed, it is evident that we can extend the construction technique of the geodetic system of Diophantine equations to any Moore graph. Let $x_1, x_2,..., x_{15}$ be the edges of the Petersen graph (see Figure 5). Let us construct its geodetic system of sixteen Diophantine equations whose unknowns are edges $x_1, x_2,..., x_{15}$ as follows:

$$\begin{aligned}
x_1+x_2+x_3+x_4+x_5 &= 2k_1+1 \\
x_1+x_5+x_7+x_{10}+x_{11} &= 2k_2+1 \\
x_1+x_2+x_6+x_8+x_{12} &= 2k_3+1 \\
x_1+x_6+x_7+x_{13}+x_{15} &= 2k_4+1 \\
x_5+x_6+x_{10}+x_{12}+x_{14} &= 2k_5+1 \\
x_4+x_5+x_6+x_9+x_{15} &= 2k_6+1 \\
x_1+x_2+x_3+x_6+x_9+x_{15} &= 2d+2 \\
x_2+x_3+x_4+x_7+x_{10}+x_{11} &= 2d+2 \quad (7)\\
x_3+x_4+x_5+x_6+x_8+x_{12} &= 2d+2 \\
x_1+x_4+x_5+x_7+x_9+x_{13} &= 2d+2 \\
x_1+x_2+x_5+x_8+x_{10}+x_{14} &= 2d+2 \\
x_1+x_6+x_7+x_{11}+x_{12}+x_{14} &= 2d+2 \\
x_2+x_7+x_8+x_{12}+x_{13}+x_{15} &= 2d+2 \\
x_3+x_8+x_9+x_{11}+x_{13}+x_{14} &= 2d+2 \\
x_4+x_9+x_{10}+x_{12}+x_{14}+x_{15} &= 2d+2 \\
x_5+x_6+x_{10}+x_{11}+x_{13}+x_{15} &= 2d+2
\end{aligned}$$

where $d \geq 2$ and $k_j$, $j = 1, 2,..., 6$ are positive integers such that $2 \leq k_j \leq d$, $j = 1, 2,...,6$. Notice that in system (7), the value of $d$ represents the diameter of a given graph $G$ homeomorphic to the Petersen graph.

As before, the unknowns of the first six equations of system (7) are the edges of the circuits belonging to a fundamental circuit basis associated to the Petersen graph. This basis is indeed composed of odd circuits of minimal length 5 with respect to a spanning tree $T = T(3, 2)$. On the other hand, the unknowns of the last ten equations of system (7) determine the edges of all circuits of length 6 contained in the Petersen graph. Notice that according to Theorem 6, the Petersen graph has twelve circuits of length 5 and ten of length 6.

**Theorem 10.** Every natural solution of the geodetic system of Diophantine equations associated with the Petersen graph determines a geodetic graph homeomorphic to it.

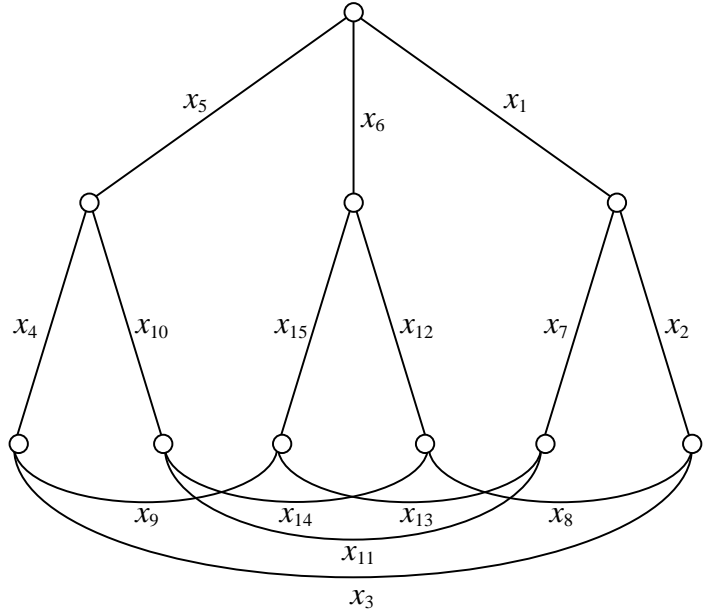

Fig. 5. The Petersen graph represented in the form of a tree $T(3, 2)$ and all edges of $S$.

**Proof.** Firstly, we will prove that the equations defining the circuits of length 5 that are not included in system (7) also determine and odd number:

(i) $(x_1+x_5+x_7+x_{10}+x_{11})+(x_1+x_2+x_6+x_8+x_{12})+(x_5+x_6+x_{10}+x_{12}+x_{14})=(2k_2+1)+(2k_3+1)+(2k_5+1)$,
$$x_2+x_7+x_8+x_{11}+x_{14} = 2(k_2+k_3+k_5-x_1-x_5-x_6-x_{10}-x_{12})+3.$$

(ii) $(x_1+x_2+x_3+x_4+x_5)+(x_1+x_2+x_6+x_8+x_{12})+(x_4+x_5+x_6+x_9+x_{15})=(2k_1+1)+(2k_3+1)+(2k_6+1)$,
$$x_3+x_8+x_9+x_{12}+x_{15} = 2(k_1+k_3+k_6-x_1-x_2-x_4-x_5-x_6)+3.$$

(iii) $(x_1+x_5+x_7+x_{10}+x_{11})+(x_1+x_6+x_7+x_{13}+x_{15})+(x_4+x_5+x_6+x_9+x_{15})=(2k_2+1)+(2k_4+1)+(2k_6+1)$
$$x_4+x_9+x_{10}+x_{11}+x_{13} = 2(k_2+k_4+k_6-x_1-x_5-x_6-x_7-x_{15})+3.$$

(iv) $(x_1+x_2+x_3+x_4+x_5)+(x_1+x_6+x_7+x_{13}+x_{15})+(x_4+x_5+x_6+x_9+x_{15})=(2k_1+1)+(2k_4+1)+(2k_6+1)$,
$$x_2+x_3+x_7+x_9+x_{13} = 2(k_1+k_4+k_6-x_1-x_4-x_5-x_6-x_{15})+3.$$

(v) $(x_1+x_2+x_3+x_4+x_5)+(x_1+x_2+x_6+x_8+x_{12})+(x_5+x_6+x_{10}+x_{12}+x_{14})=(2k_1+1)+(2k_3+1)+(2k_5+1)$,
$$x_3+x_4+x_8+x_{10}+x_{14} = 2(k_1+k_3+k_5-x_1-x_2-x_5-x_6-x_{12})+3.$$

(vi) $(x_1+x_5+x_7+x_{10}+x_{11})+2(x_1+x_2+x_6+x_8+x_{12})+(x_1+x_6+x_7+x_{13}+x_{15})+(x_5+x_6+x_{10}+x_{12}+x_{14}) =$
$= (2k_2+1)+2(2k_3+1)+(2k_4+1)+(2k_5+1)$,

$$x_{11}+x_{12}+x_{13}+x_{14}+x_{15} = 2(k_2+2k_3+k_4+k_5-2x_1-x_2-x_5-2x_6-x_7-x_8-x_{10}-x_{12})+5.$$

Assume that for fixed positive integers $d$ and $k_j$, $j = 1, 2,…,6$ system (7) has a natural solution $(x_1, x_2,…, x_{15}) = (a_1, a_2,…, a_{15})$ that determines a graph $G$ homeomorphic to the Petersen graph. Knowing that $d$ is the diameter of $G$, from the construction of system (7) it follows that $G$ does not contain an even circuit whose length is $2d$ or less. This means that the distance between any pair of vertices $u$, $v$ in $G$ is not greater than $d$ and the minimal even circuit of $G$ has length $2d + 2$ whose diameter is $d + 1$. So, $G$ does not contain even circuits $C$ in which for a pair of $C$-opposite vertices $u$, $v$ $d_G(u, v) = d(C)$. Consequently according to Lemma 1, $G$ is geodetic.

**Corollary 3.** Let $G$ be a graph homeomorphic to the Petersen graph; $G$ is geodetic if:

(1) Each path of $G$ that contains two segments is a geodesic.
(2) Each circuit of $G$ that contains exactly five segments is odd.
(3) All circuits of $G$ that contain exactly six segments have equal length.

**Remark 1:** It is clear that Corollary 3's condition (1) is equivalent to the fact that for the geodetic system of Diophantine equations associated to the Petersen graph, $d \geq 2$ and $k_j$, $j = 1, 2,…, 6$ are positive integers such that $2 \leq k_j \leq d$, $j = 1, 2,…,6$.

It is evident that the result described in Corollary 3 can be extended to any Moore graph.

**Theorem 11.** Let $G$ be a graph homeomorphic to any Moore graph of diameter $d$; $G$ is geodetic if:
(1) Each path of $G$ that contains $d$ segments is a geodesic.
(2) Each circuit of $G$ that contains exactly $2d + 1$ segments is odd.
(3) All circuits of $G$ that contain exactly $2d + 2$ segments have equal length.

**Remark 2:** Theorem 11's conditions (1), (2), (3) are, according to Theorem 2, necessary and sufficient conditions for the geodeticity of a graph $G$ homeomorphic to a complete graph $K_4$ whose diameter is 1. However, they are sufficient (but not necessary) for the geodeticity of a graph $G$ homeomorphic to a complete graph $K_n$, $n \geq 5$.

**Remark 3:** We believe that Theorem 11's conditions (1), (2), and (3) are necessary and sufficient conditions for the geodeticity of a graph $G$ homeomorphic to any Moore graph distinct of $K_n$, $n \geq 5$.

A geodetic system of Diophantine equations can be constructed for any given geodetic graph. Next, we describe the procedure of construction which is based on the results obtained in [7].

We denote the $i$-th neighborhood of vertex $v$ by

$$N_i(v) = \{u \in V(G)/d(u, v) = i\}.$$

Using this concept, the authors in [7] proved

**Theorem 12.** A graph $G$ is geodetic if and only if for every $v \in V(G)$ each vertex of $N_r(v)$ is adjacent to a unique vertex of $N_{r-1}(v)$ for each $r$ with $2 \leq r \leq d$.

According to Theorem 12, it is clear that every geodetic graph may be represented as the union of a tree $T = T(k, d)$ with levels 0, 1, 2,…, $d$ (see Figure 6) and a set $S$ of edges which suitably connect the vertices of levels 1, 2,…, $d$. Any vertex of the graph may be used as origin or topmost point.

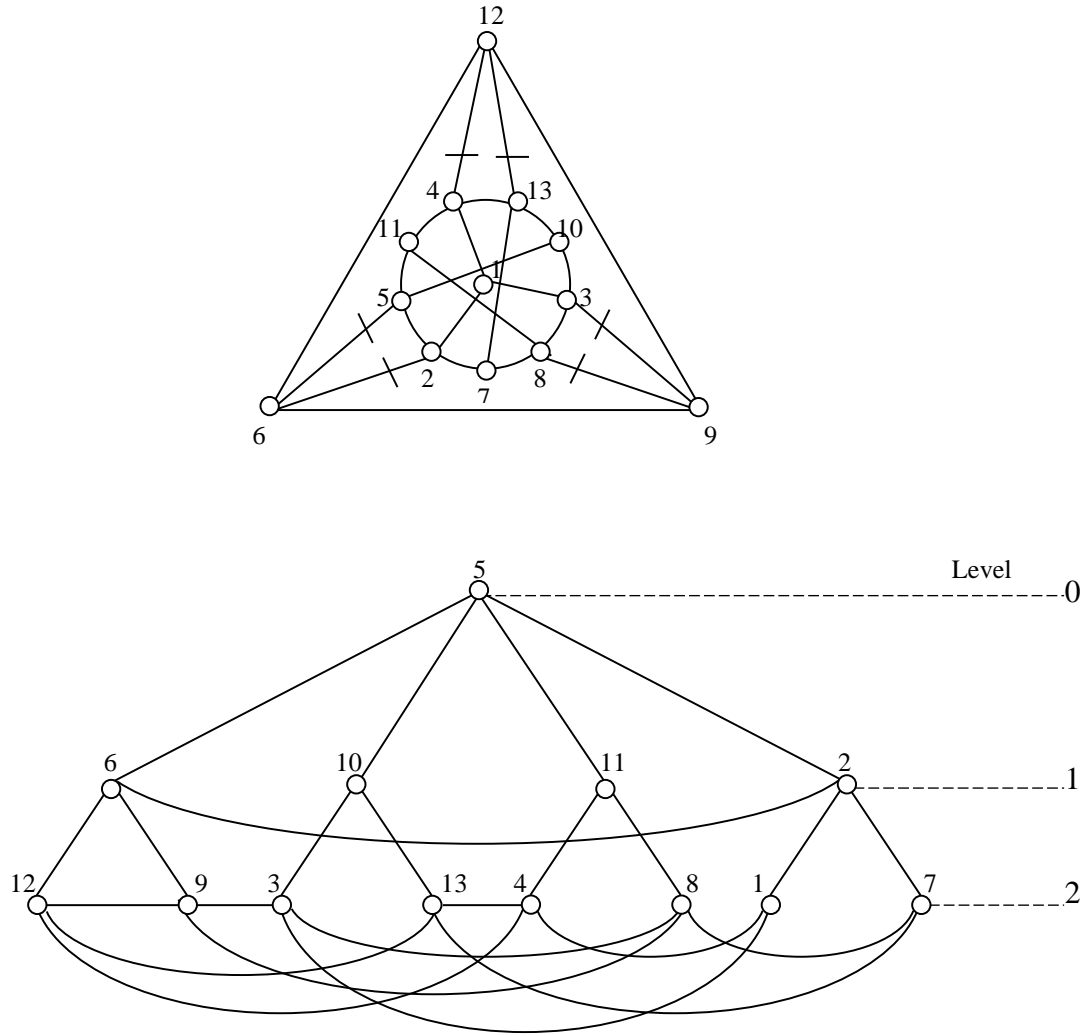

Fig. 6. A geodetic block $B(\delta, d) = B(3, 2)$ represented as the union of a tree $T$ with levels 0, 1, 2 and a set $S$ of edges which suitably connect the vertices of levels 1 and 2.

Let us construct the geodetic system of Diophantine equations associated with a geodetic graph $G$ so that its unknowns determine a fundamental circuit basis composed of $m$ odd circuits with respect to the spanning tree $T$ with levels 0, 1, 2,…, $d$ and all circuits of length 4, 6,…, $2d + 2$ contained in $G$.

We denote a geodetic block containing at least one vertex of minimal degree $\delta \geq 3$ and diameter $d \geq 2$ by $B(\delta, d)$, for which may exist more than one geodetic graph. $B(\delta, d)$ is defined as a graph that does not contain even circuits of length less than $2d + 2$ (see Figure 6).

The vector of the constant terms of the geodetic system of Diophantine equations associated with a graph $B(\delta, d)$ is written in the form:

$$\begin{pmatrix} b_1 \\ \ldots \\ \ldots \\ b_m \\ b_{m+1} \\ \ldots \\ \ldots \\ b_p \end{pmatrix} = \begin{pmatrix} 2k_1 + 1 \\ \ldots \\ \ldots \\ 2k_m + 1 \\ 2D + 2 \\ \ldots \\ \ldots \\ 2D + 2 \end{pmatrix},$$

Notice that $d$ is the diameter of the original graph $B(\delta, d)$ and $D$ is the diameter of any graph $G$ homeomorphic to $B(\delta, d)$. If this is the case, then for a fixed positive integer $D \geq d$, $k_j$, $j = 1, 2,\ldots, m$ is a positive integer such that $L \leq 2k_j + 1 \leq 2D + 1$, $j = 1, 2,\ldots, m$, where $L$ is 3 if in the system, there is an equation that determines a minimal odd circuit of length 3, $L$ is 5 if in the system, there is an equation that determines a minimal odd circuit of length 5, and so on.

**Theorem 13.** Every natural solution of the geodetic system of Diophantine equations associated with a graph $B(\delta, d)$ determines a geodetic graph homeomorphic to $B(\delta, d)$.

The proof follows immediately from Lemma 1.

A graph $G$ of diameter $D = 3$ homeomorphic to $B(\delta, d) = B(3, 2)$ is shown in Figure 6. Notice that $G$, which is geodetic, does not determine a natural solution of the geodetic system of Diophantine equations associated with $B$ within the boundaries established for the constant terms; that is, when $D = 3$, $G$ contains not only even circuits of length $2D + 2 = 2(3) + 2 = 8$, but also it contains even circuits of length $2D = 2(3) = 6$. It is evident that if we choose, in an adequate way, the values of the constant terms of the geodetic system of Diophantine equations, then the set of all its natural solutions will determine the class of all geodetic graphs homeomorphic to a given geodetic one. Unfortunately, the algorithm of choosing the boundaries for these values, which would allow us to obtain the class of all geodetic graphs homeomorphic to a given geodetic one, is unknown to us.

By itself, the construction of the geodetic system of Diophantine equations associated with a given geodetic graph does not give us precise information of how to generate the class of all geodetic graphs homeomorphic to a given geodetic one. In order to solve the problem of generating this class of graphs, we have elaborated an algorithmic procedure with the help of which it is possible to find all natural solutions of the geodetic system of Diophantine equations. In the following section, we describe an algorithm for finding all natural solutions of the geodetic system of Diophantine equations associated with the Petersen graph. In addition, the enumeration of the class of geodetic graphs homeomorphic to certain Moore graphs has been performed. This was carried out by using general foundations of enumerative combinatorics [3].

## 3. On the Enumeration of some Classes of Geodetic Graphs Homeomorphic to a Given Geodetic Graph.

Next, we will examine the problem of enumerating the class of all geodetic graphs homeomorphic to certain Moore graphs. But before, we will study some basic concepts that are necessary in our subsequent exposition. The first concept, an essential one in combinatorial theory, is called *partition*.

A partition of a positive integer $i$ is a way of representing it as the unordered sums of other positive integers in the form

$$i = x_1 + x_2 + ... + x_k, \quad x_j > 0, \quad j = 1,...,k.$$

We denote by $p_k(i)$ the number of unordered partitions of $i$ into $k$ elements. An unordered partition may be represented in standard form by writing its parts in descending order. Thus, $p_k(i)$ is the number of solutions in $x_j$ positive integers solutions of an equation

$$i = x_1 + x_2 + ... + x_k, \quad x_1 \geq x_2 \geq ... \geq x_k \geq 1.$$

It is known [3] that

$$p_k(i) = p_k(i - k) + p_{k-1}(i - k) + ... + p_1(i - k). \quad (8)$$

This recurrence relation determines the value of $p_k(i)$ with initial conditions $p_k(i) = 0$ for $i < k$ and $p_k(k) = 1$. We have used (8) to construct a table of values for $p_k(i)$ (see Table 1).

Table 1. The values of $p_k(i)$.

| k \ i | 1 | 2 | 3 | 4 | 5 | 6 | 7 | 8 | 9 | 10 | 11 | 12 | 13 | 14 | 15 |
|---|---|---|---|---|---|---|---|---|---|---|---|---|---|---|---|
| 1 | 1 | 1 | 1 | 1 | 1 | 1 | 1 | 1 | 1 | 1 | 1 | 1 | 1 | 1 | 1 |
| 2 | 0 | 1 | 1 | 2 | 2 | 3 | 3 | 4 | 4 | 5 | 5 | 6 | 6 | 7 | 7 |
| 3 | 0 | 0 | 1 | 1 | 2 | 3 | 4 | 5 | 7 | 8 | 10 | 12 | 14 | 16 | 19 |
| 4 | 0 | 0 | 0 | 1 | 1 | 2 | 3 | 5 | 6 | 9 | 11 | 15 | 18 | 23 | 27 |
| 5 | 0 | 0 | 0 | 0 | 1 | 1 | 2 | 3 | 5 | 7 | 10 | 13 | 18 | 23 | 30 |
| 6 | 0 | 0 | 0 | 0 | 0 | 1 | 1 | 2 | 3 | 5 | 7 | 11 | 14 | 20 | 26 |
| 7 | 0 | 0 | 0 | 0 | 0 | 0 | 1 | 1 | 2 | 3 | 5 | 7 | 11 | 15 | 21 |
| 8 | 0 | 0 | 0 | 0 | 0 | 0 | 0 | 1 | 1 | 2 | 3 | 5 | 7 | 11 | 15 |
| 9 | 0 | 0 | 0 | 0 | 0 | 0 | 0 | 0 | 1 | 1 | 2 | 3 | 5 | 7 | 11 |
| 10 | 0 | 0 | 0 | 0 | 0 | 0 | 0 | 0 | 0 | 1 | 1 | 2 | 3 | 5 | 7 |

Notice that after completing Table 1 by using correlation (8), the following recurrence relation also stands for a fixed integer $i > 0$:

$$p_1(i) + p_2(i) + ... + p_k(i) = p_k(i + k).$$

In this direction, recall that the number of non-negative integer solutions to $n = x_1 + x_2 + ... + x_m$ is

$$\binom{n+m-1}{m-1} \qquad (9)$$

Let *A* be a collection composed of *n* elements of *m* different types

$$A = \{a_1, a_2,..., a_n\}.$$

Assume that for elements of the same type there is no distinction. Let $r_1$ be the number of elements belonging to the first type, $r_2$ the number of elements belonging to the second type, and so on, $r_m$ the number of elements belonging to the *m-th* type, where $r_1 + r_2 + ... + r_m = n$.

The different finite collections containing *n* elements, from which $r_1$ belong to type 1, $r_2$ belong to type 2, and so on, $r_m$ belong to type *m*, are called permutations with repetition composed by $r_1, r_2, ..., r_m$.

The number of different permutations with repetition, denoted by $C_n(r_1, r_2, ..., r_m)$, is

$$C_n(r_1, r_2,..., r_m) = n!/[\; r_1!\; r_2!... \; r_m!].$$

Numbers $C_n(r_1, r_2,..., r_m)$ are called *polynomial coefficients*. These numbers are the coefficients for calculating the sum of *m* magnitudes raised to a power of *n*

$$(a_1 + a_2 +...+ a_m)^n = \sum C_n(r_1, r_2,..., r_m)\; a_1^{r_1}\; a_2^{r_2} ... \; a_m^{r_m}. \qquad (10)$$

On the right side of equation (10), we can find the sum of all possible addends of type $C_n(r_1, r_2,..., r_m)\; a_1^{r_1}\; a_2^{r_2} ... \; a_m^{r_m}$ associated with identity $r_1 + r_2 + ... + r_m = n$.

If $a_1 = a_2 = ... = a_m$, then from formula (10) it follows that

$$\sum C_n(r_1, r_2,..., r_m) = m^n. \qquad (11)$$

The following results are established for geodetic graphs homeomorphic to a complete graph $K_n$ on *n* vertices. Here, Plesnik numbers $i(u_r)$, $r = 1,...,n$ of a Plesnik graph $K_n^i$ will be denoted by $i_r$, $r = 1,...,n$.

Assume that $v_1, v_2,...,v_k$ is a labeling of the nodes of a complete graph $K_n$ so that to nodes $v_1, v_2,...,v_k$ are assigned, respectively, Plesnik numbers $i_1, i_2,...,i_k$, $i_j > 0$, $j = 1,...,k$ such that $i = i_1 + i_2 +...+ i_k$, $i_1 \geq i_2 \geq ...\geq i_k \geq 1$, $k = 1, 2,...,n$. Every pair of different partitions $i = i_1 + i_2 +...+ i_k$ generates Plesnik numbers which, once assigned to the corresponding nodes of $K_n$, determine a pair of non-isomorphic related-to-the-nodes geodetic graphs $K_n^i$ homeomorphic to $K_n$. For instance, when $i = 3$, we have three different partitions $i = 3$, $i = 2 + 1$, and $i = 1 + 1 + 1$ which determine the three non-isomorphic related-to-the-nodes geodetic graphs $K_5^3$ of Figure 2. Notice that the labeling of the nodes is here used with the only purpose of assigning to the nodes their corresponding Plesnik numbers and it is not used when determining non-isomorphic geodetic graphs homeomorphic to $K_n$. For the sake of simplicity, the number of the class of geodetic graphs this way defined will be called "the total number of non-isomorphic geodetic graphs homeomorphic to $K_n$." In this direction, we have

**Lemma 2.** Assume that for $K_n$, *n* is a fixed positive integer. Each value $p_k(i)$ in column *i* of Table 1, for which $k = 1, 2,...,n$, is the number of non-isomorphic geodetic graphs $K_n^i$ homeomorphic to a complete graph $K_n$ that have Plesnik numbers $i_1, i_2,...,i_k$, $i_j > 0$, $j = 1,...,k$ such that $i = i_1 + i_2 +...+ i_k$, $i_1 \geq i_2 \geq ... \geq i_k \geq 1$.

**Proof.** Let $k$ be a fixed positive integer, $k \leq n$. If $k > i$, then the value of $p_k(i)$ is zero and graph $K_n^i$ is not possible to construct. If $k \leq i$, then a one-to-one correspondence between $K_n$'s nodes $v_1, v_2,...,v_k$, and Plesnik numbers $i_1, i_2,...,i_k$, $i_j > 0$, $j = 1, 2,...,k$ such that $i = i_1 + i_2 +...+ i_k$, $i_1 \geq i_2 \geq ... \geq i_k \geq 1$, generates graphs $K_n^i$ that do not preserve adjacency with respect to its nodes since the values $i_1, i_2,...,i_k$, $i_j > 0$ change from partition to partition.

**Theorem 15.** Assume that $i$ and $n$ are fixed positive integers. The total number of non-isomorphic geodetic graphs $K_n^i$ homeomorphic to a complete graph $K_n$ is

$$p_1(i) + p_2(i) +...+ p_n(i) = p_n(i + n).$$

**Proof:** When $i$ and $n$ are fixed positive integers, $p_1(i)$ determines the number of different partitions of the form $i = i_1 + i_2 +...+ i_k$ for $k = 1$, $p_2(i)$ determines the number of different partitions of the form $i = i_1 + i_2 +...+ i_k$ for $k = 2$, and so on, $p_n(i)$ determines the number of different partitions of the form $i = i_1 + i_2 +...+ i_k$ for $k = n$. So, $p_1(i) + p_2(i) +...+ p_n(i)$ determines the total number of different partitions $i = i_1 + i_2 +...+ i_k$, $k = 1, 2,..., n$. According to Lemma 2, a one-to-one correspondence between $K_n$'s nodes $v_1, v_2,...,v_k$, and Plesnik numbers $i_1, i_2,...,i_k$, $i_j > 0$, $j = 1, 2,...,k$ such that $i = i_1 + i_2 +...+ i_k$, $i_1 \geq i_2 \geq ... \geq i_k \geq 1$, $k = 1, 2,..., n$ generates graphs $K_n^i$ that are non-isomorphic geodetic graphs homeomorphic to $K_n$. Therefore, $p_1(i) + p_2(i) +...+ p_n(i) = p_n(i + n)$ is the total number of non-isomorphic geodetic graphs $K_n^i$ homeomorphic to $K_n$ for $i$ and $n$ fixed positive integer numbers.

Now, assume that $v_1, v_2,...,v_n$ is a labeling of the nodes of a complete graph $K_n$ so that to nodes $v_1, v_2,...,v_n$ are assigned, respectively, Plesnik numbers $i_1, i_2,...,i_n$, $i_j \geq 0$, $j = 1,...,n$ such that $i = i_1 + i_2 +...+ i_n$. Every pair of partitions $i = i_1 + i_2 +...+ i_n$ generates Plesnik numbers which, once assigned to the corresponding nodes of $K_n$, determine a pair of non-isomorphic related-to-the-nodes labeled geodetic graphs $K_n^i$ homeomorphic to $K_n$. For instance, if we assign to the nodes of a graph $K_5$ (located to the left of Figure 2) labels $v_1, v_2,...,v_5$ and, by turns, to $v_1$ we assigned Plesnik number 3 and to the other nodes Plesnik number 0, to $v_2$ Plesnik number 3 and to the other nodes Plesnik number 0, and so on, to $v_5$ Plesnik number 3 and to the other nodes Plesnik number 0, then we will get five non-isomorphic related-to-the-nodes labeled geodetic graphs $K_n^i$ homeomorphic to $K_n$. For the sake of simplicity, the number of the class of geodetic graphs this way defined will be called "the general number of all geodetic graphs homeomorphic to $K_n$." In this new direction, we have established

**Theorem 16.** Let $i$, $n$ be fixed positive integers. The general number of all geodetic graphs $K_n^i$ homeomorphic to a complete graph $K_n$ is

$$\binom{i + n - 1}{n - 1}$$

**Proof.** We need to find the general number of solutions of $i = i_1 + i_2 +...+ i_n$ in non-negative integer numbers $i_r$, $r = 1,...,n$. According to formula (9), that number is

$$\binom{i+n-1}{n-1}$$

**Corollary 4.** For a fixed positive integer $d$ such that $k_j \leq d$, $j = 1, 2, 3$, the number of all natural solutions of the geodetic system of Diophantine equations associated with $K_4$ is

$$\binom{d+2}{3}$$

**Proof.** Since for any fixed positive integer $i$, the diameter of a geodetic graph $K_4^i$ homeomorphic to $K_4$ is $d = i + 1$, then the proof follows from Theorem 16 & Corollary 1.

**Corollary 5.** The total number of non-isomorphic geodetic graphs of diameter $d$ and the general number of all geodetic graphs of diameter $d$ homeomorphic to a complete graph $K_4$ is, respectively,

$$p_4(d+3) \text{ and } \binom{d+2}{3}$$

Next, we describe the algorithm of finding all natural solutions of the geodetic system of Diophantine equations associated with graph $B(\delta, d)$.

    **Step 1.** Assign to diameter $D$ any positive integer value $D \geq d$.
    **Step 2.** Assume that the fundamental circuit basis of graph $B(\delta, d)$ composed of odd circuits with length less than $2d + 2$ contains a system of $n_0$ odd circuits of length 3, $n_1$ odd circuits of length 5, and so on, $n_k$ odd circuits of length $2k + 3 = 2d + 1$, where $m = n_0 + n_1 + ...+ n_k$. Since there are $D$ distinct odd numbers greater than or equal to 3 and less than $2D + 2$, $D - 1$ distinct odd numbers greater than or equal to 5 and less than $2D + 2$, and so on, $D - k$ distinct odd numbers greater than or equal to $2k + 3$ and less than $2D + 2$, find all permutations with repetition of size $n_0, n_1,..., n_k$ of those $D, D - 1,..., D - k$ odd numbers corresponding to the constant terms of the subsystem that determines odd circuits of length 3, 5,..., $2k + 3$, where the number of these permutations is $D^{n_0}$, $(D - 1)^{n_1},...,(D - k)^{n_k}$, respectively.
    **Step 3.** For the given sets $A_0, A_1,..., A_k$, where $A_0$ contains $D^{n_0}$ permutations with repetition corresponding to the constant terms of the subsystem that determines the odd circuits of length 3, $A_1$ contains $(D - 1)^{n_1}$ permutations with repetition corresponding to the constant terms of the subsystem that determines circuits of length 5, and so on, $A_k$ contains $(D - k)^{n_k}$ permutations with repetition corresponding to the constant terms of the subsystem that determines the odd circuits of length $2k + 3$, find the Cartesian product of $A_0, A_1,..., A_k$, in order to obtain all possible combinations of the described permutations with repetition whose number is $D^{n_0}(D - 1)^{n_1}...(D -k)^{n_k}$.
    **Step 4.** Solve the geodetic system of Diophantine equations of graph $B = B(\delta, d)$ whose constant terms determine each of the described combinations described at the end of step 3.
    **Step 5.** Choose those solutions of the system that are natural.
    **Step 6.** Compute the number of natural solutions for a given $D$.
    **Step 7.** Assign to diameter $D$ another value and return to step 2.

Problems arising in the general theory of geodetic graphs have a great practical importance. The algorithmic aspect of these problems is of particular interest for several applications. From a practical point of view, it is essential to find, relying on such a powerful tool as a computer, all geodetic graphs of diameter $D$ homeomorphic to a given geodetic graph $G$ of diameter $d$. For this purpose, we need to have an algorithm that works effectively, and ultimately, to have the corresponding computer program.

Let us consider that all required calculations are performed in a given abstract computer. The memory of this computer consists of a boundless number of cells designated by 1, 2, ..., $n$, .... Assume we have a direct access to all its cells. Consider the problem of representing a geodetic graph $G$ in a computer's memory. Let $V(G) = \{v_1,..., v_n\}$. We use two types of representation; the first one is the representation of $G$ by its adjacency matrix, and the second one is the representation of $G$ using the list of its edges $E = \{e_1, ..., e_m\}$ where $m = |E(G)|$, $e_i \in E(G)$. In the current problem, an essential part of the initial information is, by itself, the corresponding geodetic graph. In addition, initial data should include vertex labels (or edge labels). For instance, if our problem is to find in $G$ all different circuits of a certain length, then beside its adjacency matrix, we need to provide its vertex labels (or edge labels).

At the time this research was being completed, we wrote a computer program with the help of which, we were able to solve, multiple times, the geodetic system of Diophantine equations associated with the Petersen graph. We performed something similar for the system associated with the Hoffman-Singleton graph. However, the computer program was much more complex to elaborate since it had to enable us to find 126 odd circuits of length 5 and 5250 even circuits of length 6 in order to form its 5376 equations with 175 unknowns. Unfortunately, our machine employed an enormous amount of time when finding the values of the unknowns corresponding to a single solution. This situation complicated our search of the class of all geodetic graphs homeomorphic to the Hoffman-Singleton graph.

The algorithm described below allows us to generate all geodetic graphs of fixed diameter $D \geq d$, homeomorphic to a given geodetic graph $G$ of diameter $d$.

**Step 1.** Sketch $G$ as the union of its spanning tree $T$ with levels 0, 1, 2,…,$d$ and a set $S$ of edges that suitably connect the vertices of levels 1, 2,…, $d$ by adding, one by one, the edges of $S$ to $T$. Then find a fundamental circuit basis composed of odd circuits with respect to $T$.

**Step 2.** Determine all circuits of length 4, 6,..., $2d + 2$ contained in $G$.

**Step 3.** Construct the geodetic system of Diophantine equations whose unknowns determine all circuits of steps 1 and 2.

**Step 4.** Apply the required algorithm to find all possible values that may be assigned to its vector of constant terms.

**Step 5.** Solve the system for each set of values established in step 4.

**Step 6.** Choose the solutions of the system that are natural.

**Step 7.** Represent all geodetic graphs homeomorphic to $G$ using the natural solutions found in step 6.

**Step 8.** Assign another value to $D$ and return to step 4.

Now, we will be able to apply the whole set of results established above to a specific geodetic graph. We refer to the Petersen graph. As we will note later, the Petersen graph represents a sort of "model graph" for which it can be shown that the geodetic system of Diophantine equations and the algorithm of enumeration of its natural solutions work well on the solution of the problem of finding the class of all geodetic graphs homeomorphic to a given geodetic one.

**Theorem 17.** For a fixed positive integer $d \geq 2$ such that $2 \leq k_j \leq d$, $j = 1, 2,\ldots, 6$, the general number of natural solutions of the geodetic system of Diophantine equations associated with the Petersen graph does not exceed $(d - 1)^6$.

**Proof.** For $d \geq 2$ such that $2 \leq k_j \leq d$, $j = 1, 2,\ldots, 6$, there exist $d - 1$ different odd numbers greater than or equal to 5 and less than $2d + 2$. Therefore, the constant terms of the first six equations of system (7) form a collection of odd numbers composed of $n = 6$ elements belonging to a set of $m = d - 1$ different odd numbers. According to formula (11), the number of all of these collections is $m^n = (d - 1)^6$.

Table 2 shows all possible collections $A = \{2k_1 + 1, 2k_2 + 1,\ldots,2k_6 + 1\}$ of the constant terms of the first six equations of system (7) and their number of permutations with repetition when $d = 7$. With the help of the set of collections $A$ shown in Table 2, it is possible to find all other collections $A$ and their number of permutations with repetition not only for $d = 7$, but also for any other value of $d$.

Notice that for any $d \geq 2$, the number of permutations with repetition of all collections $A$ belonging to groups I, II, III, IV, V, VI is, respectively,

$$1\binom{d-1}{1}, \quad 62\binom{d-1}{2}, \quad 540\binom{d-1}{3}, \quad 1560\binom{d-1}{4}, \quad 1800\binom{d-1}{5}, \quad 720\binom{d-1}{6}.$$

Now, we are ready to describe the algorithm of enumerating all natural solutions of the geodetic system of Diophantine equations associated with the Petersen graph.

**Step 1.** Assign to diameter $d$ a given value for any positive integer $d \geq 2$.

**First Iteration.**

**Step 2.** Find all combinations without repetition of $d - 1$ different odd numbers greater than or equal to 5 and less than $2d + 2$ taken 1 at a time whose number is $\binom{d-1}{1}$.

**Step 3.** For the first of the combinations of Step 2, find a collection $A(\mathrm{I}) = \{2k_1 + 1, 2k_2 + 1,\ldots, 2k_6 + 1\}$ belonging to group I (see Table 2) and repeat the procedure with each of the next combinations until $1 \cdot \binom{d-1}{1}$ of these collections are obtained.

**Step 4.** For the first of the collections of Step 3, find all $C_6$ permutations with repetition and repeat the procedure with each of the next collections until $1 \cdot \binom{d-1}{1}$ of these permutations are obtained. We denote them by $A_i(\mathrm{I}) = \{2k_1 + 1, 2k_2 + 1,\ldots,2k_6 + 1\}$, $i = 1,\ldots, \binom{d-1}{1}$.

**Step 5.** Choose the first permutation $A_i(I) = A_1(I)$.
**Step 6.** Solve system (7) for the corresponding values of this permutation.
**Step 7.** If system (7) has a natural solution, then go to Step 8. Otherwise go to Step 9.
**Step 8.** Print the natural solution.
**Step 9.** Choose permutation $A_i(I) = A_{i+1}(I)$ and go to Step 6.

Table 2. A list of all collections $A$ of the constant terms of the first six equations of system (7) for $d = 7$.

| Group | A | $C_6$ | $\Sigma C_6$ | Number of permutations per group |
|---|---|---|---|---|
| I | 5 5 5 5 5 5 | 1 | 1 | 6 |
| II | 5 7 7 7 7 7 | 6 | 62 | 930 |
| | 5 5 7 7 7 7 | 15 | | |
| | 5 5 5 7 7 7 | 20 | | |
| | 5 5 5 5 7 7 | 15 | | |
| | 5 5 5 5 5 7 | 6 | | |
| III | 5 7 9 9 9 9 | 30 | 540 | 10800 |
| | 5 7 7 9 9 9 | 60 | | |
| | 5 7 7 7 9 9 | 60 | | |
| | 5 7 7 7 7 9 | 30 | | |
| | 5 5 7 9 9 9 | 60 | | |
| | 5 5 7 7 9 9 | 90 | | |
| | 5 5 7 7 7 9 | 60 | | |
| | 5 5 5 7 9 9 | 60 | | |
| | 5 5 5 7 7 9 | 60 | | |
| | 5 5 5 5 7 9 | 30 | | |
| IV | 5 7 9 11 11 11 | 120 | 1560 | 23400 |
| | 5 7 9  9 11 11 | 180 | | |
| | 5 7 9  9  9 11 | 120 | | |
| | 5 7 7  9 11 11 | 180 | | |
| | 5 7 7  9  9 11 | 180 | | |
| | 5 7 7  7  9 11 | 120 | | |
| | 5 5 7  9 11 11 | 180 | | |
| | 5 5 7  9  9 11 | 180 | | |
| | 5 5 7  7  9 11 | 180 | | |
| | 5 5 5  7  9 11 | 120 | | |
| V | 5 7 9 11 13 13 | 360 | 1800 | 10800 |
| | 5 7 9 11 11 13 | 360 | | |
| | 5 7 9  9 11 13 | 360 | | |
| | 5 7 7  9 11 13 | 360 | | |
| | 5 5 7  9 11 13 | 360 | | |
| VI | 5 7 9 11 13 15 | 720 | 720 | 720 |

**Second Iteration.**

**Step 2.** Find all combinations without repetition of $d - 1$ different odd numbers greater than or equal to 5 and less than $2d + 2$ taken 2 at a time whose number is $\binom{d-1}{2}$.

**Step 3.** For the first of the combinations of Step 2, find 5 collections $A(\text{II}) = \{2k_1 + 1, 2k_2 + 1,..., 2k_6 + 1\}$ belonging to group II (see Table 2) and repeat the procedure with each of the next combinations until $5 \cdot \binom{d-1}{1}$ of these collections are obtained.

**Step 4.** For the first of the collections of Step 3, find all $C_6$ permutations with repetition and repeat the procedure with each of the next collections until $62 \cdot \binom{d-1}{2}$ of these permutations are obtained. We denote them by $A_i(\text{II}) = \{2k_1 + 1, 2k_2 + 1,..., 2k_6 + 1\}$, $i = 1,..., 62 \cdot \binom{d-1}{2}$.

**Step 5.** Choose the first permutation $A_i(\text{II}) = A_1(\text{II})$.
**Step 6.** Solve system (7) for the corresponding values of this permutation.
**Step 7.** If system (7) has a natural solution, then go to Step 8. Otherwise go to Step 9.
**Step 8.** Print the natural solution.
**Step 9.** Choose permutation $A_i(\text{II}) = A_{i+1}(\text{II})$ and go to Step 6.

This procedure should be continued until the sixth iteration is completed since according to Table 2, there are six groups of collections $A$. The described procedure represents the general algorithm for finding all natural solutions of the geodetic system of Diophantine equations associated with the Petersen graph.

We applied the above algorithm to find the set of geodetic graphs homeomorphic to the Petersen graph for $d = 2, 3, 4, 5, 6, 7$ (see Tables 3 through 8).

Table 3. Geodetic graphs of diameter 2 homeomorphic to the Petersen graph.

| Gr. | A | $C_6$ | $\Sigma\, C_6$ | $(d, g)$ | Graph |
|---|---|---|---|---|---|
| I | 5 5 5 5 5 5 | 1 | 1 | (2, 5) | 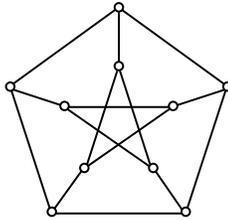 |

Table 4. Geodetic graphs of diameter 3 homeomorphic to the Petersen graph.

| Gr. | A | $C_6$ | $\Sigma\, C_6$ | $(d, g)$ | Graph |
|---|---|---|---|---|---|
| I | 5 7 7 7 7 7 | 6 | 6 | (3, 5) | 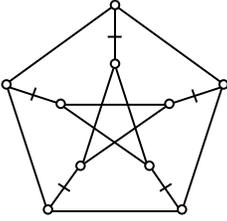 |

Table 5. Geodetic graphs of diameter 4 homeomorphic to the Petersen graph.

| Gr. | A | $C_6$ | $\Sigma C_6$ | $(d, g)$ | Graph |
|---|---|---|---|---|---|
| II | 5 9 9 9 9 9 | 6 | 21 | (4, 5) | 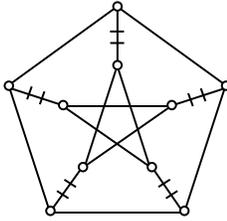 |
|    | 7 7 9 9 9 9 | 15 |    | (4, 7) | 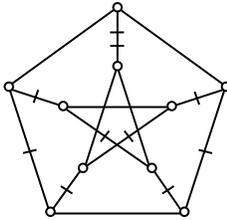 |

Table 6. Geodetic graphs of diameter 5 homeomorphic to the Petersen graph.

| Gr. | $A$ | $C_6$ | $\Sigma\, C_6$ | $(d, g)$ | Graph |
|---|---|---|---|---|---|
| II | 5 11 11 11 11 11 | 6 | 26 | (5, 5) | 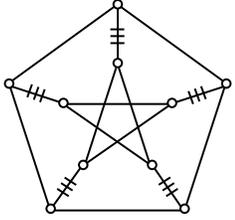 |
|  | 9 9 9 11 11 11 | 20 |  | (5, 9) | 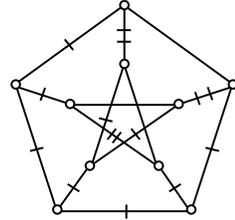 |
| III | 7 9 11 11 11 11 | 30 | 30 | (5, 7) | 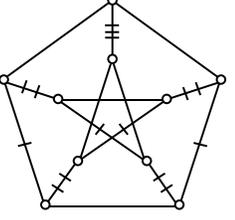 |

Table 7. Geodetic graphs of diameter 6 homeomorphic to the Petersen graph.

| Gr. | $A$ | $C_6$ | $\Sigma\,C_6$ | $(d, g)$ | Graph |
|---|---|---|---|---|---|
| II | 5 13 13 13 13 13 | 6 | 36 | (6, 5) | |
| | 9 9 13 13 13 13 | 15 | | (6, 9) | |
| | 11 11 11 11 13 13 | 15 | | (6, 11) | |
| III | 7 11 13 13 13 13 | 30 | 90 | (6, 7) | |
| | 9 11 11 13 13 13 | 60 | | (6, 9) | |

Table 8.  Geodetic graphs of diameter 7 homeomorphic to the Petersen graph.

| Gr. | $A$ | $C_6$ | $\Sigma\, C_6$ | $(d, g)$ | Graph |
|---|---|---|---|---|---|
| II | 5 15 15 15 15 15 | 6 | 12 | (7, 5) | |
|    | 13 13 13 13 13 15 | 6 |    | (7, 13) | |

Table 8. The other geodetic graphs of diameter 7 homeomorphic to the Petersen graph.

| Gr. | A | $C_6$ | $\Sigma C_6$ | $(d, g)$ | Graph |
|---|---|---|---|---|---|
| III | 7 13 15 15 15 15 | 30 | 240 | (7, 7) | |
| | 9 11 15 15 15 15 | 30 | | (7, 9) | |
| | 9 13 13 15 15 15 | 60 | | (7, 9) | |
| | 11 13 13 13 15 15 | 60 | | (7, 11) | |
| | 11 11 13 15 15 15 | 60 | | (7, 11) | |

Table 9. Geodetic graphs homeomorphic to the Petersen graph belonging to groups IV, V, VI.

| Gr. | A | (d, g) | Graph |
|---|---|---|---|
| IV | 15 17 19 19 19 21 | (10, 15) | 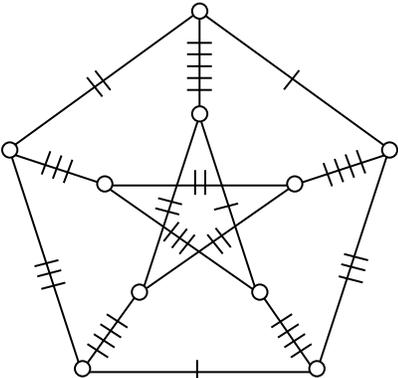 |
| V | 19 21 23 25 25 27 | (13, 19) | 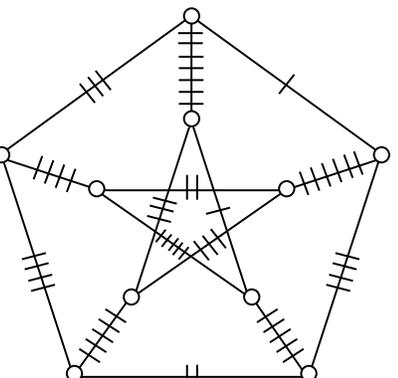 |
| VI | 25 27 29 31 33 35 | (17, 25) | 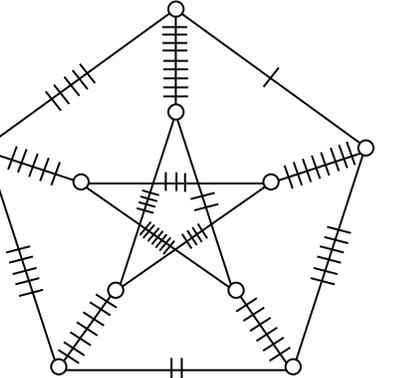 |

## 4. Conclusions

According to Tables 3 through 8, for diameters $d = 2, 3, 4, 5, 6, 7$, the total number of non-isomorphic geodetic graphs and the general number of all geodetic graphs homeomorphic to the Petersen graph is, respectively, 1, 1, 2, 3, 5, 7 and 1, 6, 21, 56, 126, 252.

**Remark 4:** If the statement in Remark 3 turns out to be true, then the following statement could also be true. The total number of non-isomorphic geodetic graphs and the general number of all geodetic graphs of diameter $d$ homeomorphic to the Petersen graph is, respectively,

$$p_6(d + 4) \text{ and } \binom{d + 3}{5}$$

Parthasarathy and Srinivasan [8] conjectured that in any geodetic graph of diameter $d$, there exist circuits of length $2d + 1$, denoted by $C_{2d + 1}$. Figure 7 exhibits a geodetic block of diameter 11 that does not contain circuits $C_{23}$.

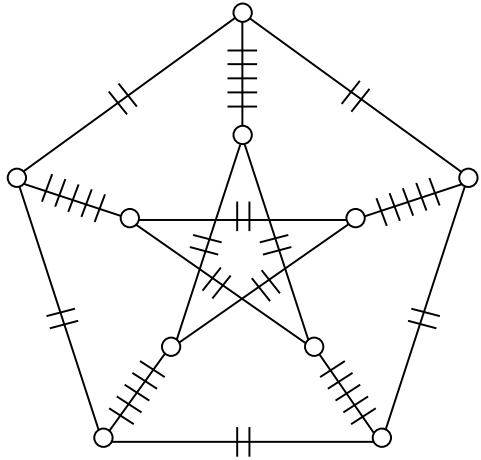

Fig. 7. A geodetic block of diameter 11 that does not contain circuits $C_{23}$.

We have solved the problem posed by Parthasarathy and Srinivasan [7] on the existence of geodetic graphs for the given values $(d, g) = (6, 11)$ (see Table 7) and $(8, 13)$ (see Figure 8).

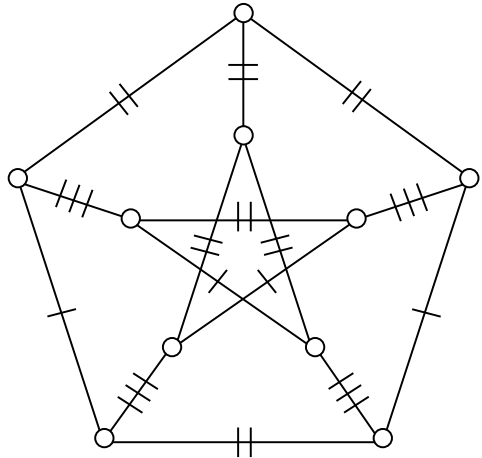

Fig. 8. A geodetic block of diameter 8 and girth 13.

Lastly, the problem of generating all geodetic graphs homeomorphic to the Petersen graph and, in general, graphs homeomorphic to Moore graphs different of a complete graph $K_n$, $n \geq 5$, which was solved above, represents a partial solution to problem 3 posed by Parthasarathy and Srinivasan in [7] on the existence and construction of geodetic blocks of diameter $d$ and girth $g \leq 2d + 1$.

**Remark 5:** It is evident that there exist geodetic graphs homeomorphic to the Petersen graph belonging to groups IV, V, VI (see Table 9).

**References.**